\documentstyle[aps,preprint]{revtex}
\tightenlines
\newcommand{\be}{\begin{equation}}
\newcommand{\ee}{\end{equation}}
\newcommand{\bear}{\begin{eqnarray}}
\newcommand{\ear}{\end{eqnarray}}
\newcommand{\ba}{\begin{eqnarray*}}
\newcommand{\ea}{\end{eqnarray*}}

\def\Journal#1#2#3#4{{#1} {\bf #2}, #3 (#4)}
\def\JPA{\em J. Phys. A}
\def\PRD{\em Phys. Rev. D}
\def\CMP{\em Commun. Math. Prys.}

\def\PRL{\em Phys. Rev. Lett.}
\def\AJ{\em Ap. J.}

\def\AsJ{\em Astrophys. J.}

\def\PR{\em Phys. Reports}

\title{Coulomb and quantum oscillator problems in conical spaces with arbitrary dimensions}
\author{J. L. A. Coelho and   R. L. P. G. Amaral \\
\small
{ \it Instituto de F\'{\i}sica - Universidade Federal Fluminense}\\
\small
{\it Av. Litor\^anea, S/N, Boa Viagem, Niter\'oi,
CEP.24210-340}\\
\small{\it Rio de Janeiro - Brasil}}

\date{\today}

\begin{document}

\maketitle

\begin{abstract}

The Schr\"odinger equations for the Coulomb and the  Harmonic oscillator potentials are solved
in the cosmic-string  conical space-time. The spherical harmonics with angular deficit are introduced.
 The  algebraic construction of the 
 harmonic oscillator eigenfunctions is performed through the introduction of non-local
ladder operators.  By exploiting the hidden symmetry of the two-dimensional harmonic oscillator the 
eigenvalues for the angular momentum operators in three dimensions  are reproduced.
 A generalization for N-dimensions is performed 
for both Coulomb and harmonic oscillator problems in angular deficit space-times.
 It is thus  established  the connection among the states and energies of both problems in these 
topologically non-trivial space-times.  

\

\end{abstract}
\vspace{0.5cm}

\section{Introduction}

There has been a growing interest for space-times with nontrivial
topology and how this can affect some aspects in classical or quantum
cosmological models as well as in quantum mechanics. This nontrivial
topology is generated
by topological defects like monopoles, strings, domain walls and branes.
Their formation are associated with phase transitions in the early universe
where the vacuum is degenerated \cite{TWBK}. Nevertheless, stable domain
walls and monopoles are disastrous for cosmological models \cite{JPP}.
However, strings cause no harm and can be a good candidate to produce
several phenomena observed in the last decades like gravitational lenses
\cite{AV,EMH}, particle production \cite{TV} and microwave sky anisotropy
\cite{AS}. The most interesting topic for our study is the metric structure
of the cosmic string space-time. The metric leads to a conic space-time.
Its locally flat geometry affects non-relativistic systems only through
the non-trivial topology. Thus, a non-relativistic particle
placed in the surroundings of a straight, infinite and static string 
will not suffer attraction by the  cosmic string gravitational field 
\cite{AV}.

For a cosmic string space-time the metric tensor in cylindrical
coordinates $(\rho ,z,\phi)$ is  $ g_{\mu\nu} = diag( 1,-1,-1,-\rho^{2})$
where $ 0 \le \rho \le \infty $, $ -\infty  \le z  \le \infty $,
$ -\pi\alpha \le \phi  \le \pi\alpha $  and $\alpha = 1-4G\mu$, with $\mu$
being the linear density of the cosmic string. The Laplace-Beltrami operator
in these coordinates

\begin{equation} \label{Eq1} \Box  \psi = \frac{1}{\sqrt{-g}}
\frac{\partial}{\partial x^\mu}\left(\sqrt{-g}g^{\mu\nu}
\frac{\partial}{\partial x^\nu}\right) \psi \end{equation}
\noindent take the same form as in the flat space-time.
When one  takes the non-relativistic limit of a system with the 
dynamics  described by  a relativistic
quantum equation  such  as the Dirac or Klein-Gordon
equations one will be led to the usual Shcr\"odinger equation but with
non-trivial boundary conditions

\begin{equation} \label{Eq2}\Phi (\phi_{0}) =
\Phi (\phi_{0} + 2\pi\alpha).
\end{equation}

\noindent The  cosmic   deficit angle
$ \delta =2\pi (1-\alpha) $ shall affect any quantum wave solution
that significantly encircles the string.

It has been noted that the cosmic string space-time affects the
quantum solutions of central potentials. The Coulomb potential has been
considered by \cite{GRV} in the context of a two-dimensional potential as
generated by the cosmic string itself.

Here we consider a general radial problem and define the spherical
harmonics taking into account the angular deficit. We apply the
results to 
the (3+1)-D Coulomb and harmonic oscillator problems obtaining the
spectra for these potentials. These spectra have been independently obtained
in \cite{VG}. We consider also a  
generalization of these problems to a $(N+1)$-dimensional space-time with 
conic topological structure. These can be originated by a $(N-2)$-brane 
of cosmic character. The central potential is added with the origin 
attached to a point of the brane. The well known relationship between 
oscillator and coulomb problems 
\cite{KM} is then generalized to hyper-conic space-times.

The structure of the paper is as follows. In section {\bf II} the solution
of the potential problems in (3+1)-Dimensions is addressed. The
spherical harmonics with angular deficit are constructed. The complete
eigenfunctions for the Coulomb and oscillator problems are presented and
the ladder operators for the latter potential introduced.  The
hidden rotational symmetry discussed. In section
{\bf III} the (N+1) dimensional generalization is performed. The
dependency of the quadratic angular momentum Casimir operator eigenvalue
on the angular deficit is obtained. The spectra of the Coulomb and
oscillator problems are obtained and related. The section {\bf IV}
presents final comments.

\section{Coulomb and  quantum oscillator problems in (3+1)-D conical
space-time}

Let us consider spherical
coordinates $(r,\theta,\phi)$ in which  the cosmic string metric tensor reads
$ g_{\mu\nu} = diag(1,-1,-r^{2},-r^{2}sin^{2}\theta)$ and
the Schr\"{o}dinger equation is written in standard form

\begin{equation}  \label{Eq3}\left[\nabla_{r}^{2} + {2 \mu
\over \hbar^{2}}V({\bf r}) + {2\mu E \over \hbar^{2}}\right]
\psi(\textbf{r}) - {\emph{\textbf{L}}^{2}
\over{\hbar^{2}r^{2}}}\psi(\textbf{r}) =0, \end{equation}

\noindent where $ {\emph{\textbf{L}}^{2}} = - \hbar^{2}
\left[{1\over{sin \theta }}{d\over d\theta}(sin \theta {d\over d\theta}) +
 {1\over{sin^{2} \theta }}{d^{2}\over
{d\Phi^{2}}}\right] $ is the angular momentum operator.

\subsection{ Eigenfunctions and energy spectrum }

Let us perform the separation of variables expressing

\begin{equation} \label{Eq4} \psi({\bf{\vec{r}}})=R(r)Y(\theta,\phi).
\end{equation}

\noindent Substituting in equation (\ref{Eq3}) we obtain the set of equations

\begin{equation} \label{Sis1}
and \hspace{2cm} \begin{array}{ll}
\left[r\frac{d^2 }{dr^2}r - \frac{2\mu}{\hbar^2}V(r) +
\frac{2\mu E}{\hbar^2}r^2\right]R(r)=\ell(\ell+1)R(r) & \\
{\emph{\textbf{L}}^{2}} Y(\theta,\phi)=\hbar^2\ell(\ell+1)
Y(\theta,\phi), &
\end{array}
\end{equation}

\noindent where $\ell(\ell +1)$ is to be specified later.
Separating the second equation with $ Y(\theta ,\phi ) = \Theta(\theta )\Phi(\phi )$
leads to

\begin{equation} \label{Sis2}
and \hspace{2cm}\begin{array}{ll}
{\frac{1}{\sin\theta}}\frac{d}{d\theta}\sin\theta\frac{d}{d\theta}\Theta
(\theta)=\lambda^{2}\Theta(\theta) & \\
\frac{d^2}{d \theta^2}\Phi(\phi)=-\lambda^{2} \Phi (\phi), &
\end{array}
\end{equation}

\noindent where $\lambda$ is another parameter to be  determined.

\subsubsection{Spherical harmonics}

\noindent The presence of the string is displayed in the azimuthal equation
which is subjected to the periodic boundary
conditions (\ref{Eq2}). So, we have

\begin{equation} \label{Eq5}\Phi_{\alpha}^{m}(\phi)=
\frac{1}{\sqrt{2\pi\alpha}} e^{i\frac{m}{\alpha} \phi}, \end{equation}

\noindent where $ m = 0,\pm 1, \pm 2... $, $\alpha \in \Re^{+}$
and we identified $ \lambda = { \frac{|m|}{\alpha} } $. Requiring regular solutions
in $\theta = 0$ we obtain $\ell = k + \frac{|m|}{\alpha}$  and the  polar solutions

\begin{equation} \label{Eq6} \Theta_{k}^{\frac{|m|}{\alpha}}
(\theta)=\sqrt{\frac{(2k+2{\frac{|m|}{\alpha}}+1)\Gamma(k+1)}
{2\Gamma(k+2{\frac{|m|}{\alpha}}+1)}}{\sin^{\frac{|m|}{\alpha}}
\theta}T_{k}^{\frac{|m|}{\alpha}}(u), \end{equation}

\noindent where $ k = 0,1,2... $, $u =cos\theta$ and $T_{k}^{\frac{|m|}{\alpha}}(u)$
are the Gegenbauer polynomials. Therefore, the generalization of the spherical
harmonics required by the cosmic string space-time are

\begin{equation}  \label{Eq7} Y_{\ell}^{\frac{m}{\alpha}}(\theta,\phi) =
{\sqrt{\frac{(2\ell+1)\Gamma(\ell -\frac{|m|}{\alpha}+1)}{4\pi\alpha
\Gamma(\ell+{\frac{|m|}{\alpha}}+1)}}}{\sin^{\frac{|m|}{\alpha}}\theta}
T_{\ell-\frac{|m|}{\alpha}}^{\frac{|m|}{\alpha}}(\cos\theta) e^{i\frac{m}{\alpha} \phi}.
\end{equation}

\noindent For the eigenvalues of quadratic Casimir operator we observe a
dependence on two integers and on the angular deficit $\alpha$

\begin{equation} \label{Eq.8} \ell(\ell+1) = \left(k + \frac{|m|}{\alpha}\right)
\left(k + \frac{|m|}{\alpha}+1\right) .\end{equation}

\noindent This dependency  of the $l$ value on each of the states within a specific
angular momentum multiplet can be understood since the operators ${\bf L}_\pm=
{\bf L}_x\pm {\bf L}_y$ are not operators that act within the Hilbert space
functions. They would create states with wrong periodicity conditions. The algebraic construction
 of the angular momentum states
is spoiled from the beggining. We will
argue latter that an attempt to redefine these operators to act in the
Hilbert space as $({\bf L}_\pm)^{\frac{1}{\alpha}}$ will not work.

\subsubsection{Radial equations}

The above procedure is valid for any problem with a radial potential
$V(r)$. Now we choose a particular potential to solve the radial equation

\begin{equation} \label{Eq9} \left[\frac{d^2 }{dr^2}+\frac{2}{r}
\frac{d}{dr} - \frac{2\mu}{\hbar^2} V({\bf r}) +\frac{2\mu E}{\hbar^2}
\right]R(r)= \left(k + \frac{|m|}{\alpha}\right)\left(k + \frac{|m|}{\alpha}+1\right)
R(r). \end{equation}

\subsubsection*{Coulomb potential}

The Coulomb potential is expressed by $ \frac{-e^2}{r}$.  Substituting
this in Ed. (13) we have

\begin{equation} \label{Eq10} \left[\frac{d^2 }{d\rho^2}+\frac{2}{\rho}
\frac{d}{d\rho} - \frac{\ell(\ell+1)}{\rho^2}+\frac{\beta}{\rho}- 
\frac{1}{4}\right]R(\rho)= 0, \end{equation} 

\noindent where $\beta$ is an arbitrary parameter, $ \rho = \frac{r}
{\beta r_{0}}$, 
$\ell = k + \frac{|m|}{\alpha}$, $E = -\frac{\epsilon_{0}}{\beta^{2}}$; 
with $ r_{0} =\frac{\hbar^2}{2\mu e^2}$ (Bohr radius divided by two) and 
$ \epsilon_{0} = \frac{\mu e^4}{2\hbar^{2}}$ (ionization energy). With the ansatz

\begin{equation} \label{Eq11} R(\rho) = C e^{(-\frac{\rho}{2})}\rho^{\ell} 
g(\rho), \end{equation}

\noindent we are led to the radial equation for $g(\rho)$

\begin{equation} \label{Eq12} \rho \frac{d^2}{d\rho^2}g(\rho)+(2\ell+2-\rho)
\frac{d}{d\rho}g(\rho) + (\beta-\ell-1)g(\rho)=0. \end{equation}

\noindent This is the associated Laguerre equation whose normalizable solutions are 
the associated Laguerre polynomials $ L^{2\ell+1}_{j}
(\rho) = \frac{\Gamma(2\ell + j + 2)}{\Gamma(j+1)}\frac{e^{\rho}}
{\rho^{2\ell+1}}\frac{d^{j}}{d\rho^{j}}[\rho^{2\ell+j+1}e^{-\rho}] $, 
with $ j= 0,1,2...$  Therefore, the solution for the radial differential equation is 

\begin{equation} \label{Eq13} R(r)= C^{\alpha}_{j,k,m}\left(\frac{1}{\beta} 
\frac{r}{r_{0}}\right)^{k+\frac{m}{\alpha}}e^{-\frac{1}{\beta}
\frac{r}{r_{0}}} L^{2k+2\frac{m}{\alpha}+1}_{j}\left(\frac{1}{\beta}\frac{r}
{r_{0}}\right), \end{equation}

\noindent where $\beta$ is given by  $ j + k + 
\frac{|m|}{\alpha}$ and $ C^{\alpha}_{j,n,m} $ is a normalization constant obtained as

\begin{equation} \label{Eq14} C^{\alpha}_{j,k,m} = \frac{1}{(j+k+\frac{m}
{\alpha})}\sqrt{\frac{\Gamma (j)}{2r_{0}^{3}\left(\Gamma (j+2k+2\frac{m}
{\alpha}+1)\right)^{3}}}. \end{equation}

\noindent For the energy spectrum we obtain

\begin{equation} \label{Eq15} E^{\alpha}_{j,k,m} = - \frac{\epsilon_{0}}
{\left(j+k+\frac{|m|}{\alpha}\right)^{2}}. \end{equation}

\noindent Clearly, the essential degeneracy is broken, but there is still  an
 accidental 
degeneracy associated with a full symmetry of the potential \cite{JPE}. 
It is important to point out  that the energy levels depend explicitly on
the angular 
deficit $\alpha$ which characterizes the global structure of the metric.

\subsubsection*{Quantum harmonic oscillator}

Proceeding as done before to the hydrogen atom, we can solve the radial 
equation (\ref{Eq10}) for $ V({{\bf{r}}}) = \frac{1}{2}\mu \omega^2 r^2 $ 
given by 

\begin{equation} \label{Eq16} \left[\frac{d^2 }{dr^2}+\frac{2}{r}\frac{d}
{dr} - \frac{\mu^2}{\hbar^2}  \omega^2 r^2 +\frac{2\mu E}{\hbar^2}\right]
R(r)= \left(k + \frac{|m|}{\alpha}\right)\left(k + \frac{|m|}{\alpha}+1\right)R(r). \end{equation}

\noindent Making the ansatz

\begin{equation} R(x) = \label{Eq17} C x^{\ell}e^{-\frac{1}{2}x^{2}} h (x),
\end{equation}

\noindent where $ x = (\frac{r}{r_{0}})^2 $ and $ r_{0} = \sqrt{\frac{\hbar}
{\mu\omega}}$. And using eq (\ref{Eq17}) in (\ref{Eq16}) we have

\begin{equation} \label{Eq18}  \frac{d^2}{dx^2}h(x)+
\left(\frac{2\ell+2}{x} -2x\right)\frac{d}{dx}h(x) + \left(\frac{2E}{\hbar\omega}-2\ell-3\right)
h(x)=0. \end{equation}

\noindent Making a new change in variable $ x \to x' = x^{\frac{1}{2}} $,
 we obtain the same equation  as in hydrogen atom (\ref{Eq12}) with 
solution $h(x^{\frac{1}{2}}) = L^{\ell +\frac{1}{2}}_{j}(x)$. Therefore, 
the solution for the radial function $ R(r)$ is given by

\begin{equation} \label{Eq19} R(r) = C^{\alpha}_{j,k,m} \left( {\frac{r}
{r_{0}}}\right )^{k+\frac{|m|}{\alpha}}e^{-\frac{1}{2}(\frac{r}{r_{0}})^{2}} 
L^{k+\frac{|m|}{\alpha}+\frac{1}{2}}_{j}
\left (\frac{r^2}{r_{0}^{2}}\right),  \end{equation}

\noindent and the normalization constant is

\begin{equation} \label{Eq20} C^{\alpha}_{j,k,m} = \sqrt{\frac{2}{r_{0}^3}
\frac{\Gamma(j+1)}{[\Gamma(j+k+\frac{|m|}{\alpha}+\frac{3}{2})]}^{3}}.
 \end{equation}

\noindent For the energy spectrum we obtain 

\begin{equation} \label{Eq21} E^{\alpha}_{j,k,m} =\hbar\omega (2j+k+
\frac{|m|}{\alpha} + \frac{3}{2}) \end{equation}

\noindent Again, there is a dependence on deficit angle and the degeneracy attributed 
to symmetry rotations (essential degeneracy) is broken. But we see clearly a 
persistence of the accidental degeneracy related with even $k$ states for $m = 0$.

\subsubsection*{Creation and Destruction operators}

The last section results show that the eigenvalues for the Harmonic 
Oscillator increase in intervals of $\hbar \omega$ or of $\hbar \omega
\alpha$. This  suggests to investigate the construction of ladder operators 
for the Harmonic Oscillator which shall produce these changes. Since the Hilbert space can be 
constructed as the tensor product ${\mathcal{E}_{\rho,\phi}}
\otimes{\mathcal{E}}_{z}$, it is sufficient to consider the 2-D 
quantum harmonic oscillator with angular deficit.

The Hamiltonian for 2-D quantum harmonic oscillator $ V(\rho) = \frac{1}{2}
\mu \omega^2\rho^2 $ without angular deficit  can be written in terms 
of creation and destruction operators of right and left ``circular quantum'' 

\begin{equation} \label{Eq22} {\bf H} = \hbar\omega({\bf a}_{d}^\dagger 
{\bf a}_{d} + {\bf a}_{g}^\dagger {\bf a}_{g} + 1), \end{equation}

\noindent these operators are defined in terms of usual $a_{x}$ and $a_{y}$ 
operators by

\begin{equation} \label{Sis3}
and \hspace{2cm} \begin{array}{ll}
a_{d} = \frac{1}{\sqrt{2}}(a_{x}-ia_{y})&  \\
a_{g} = \frac{1}{\sqrt{2}}(a_{x}+ia_{y}),& \\
\end{array}
\end{equation}

\noindent The  non-zero commutators between the four operators $ a_{d}$, 
$a_{g}$, $a_{d}^{\dagger}$, and $a_{g}^{\dagger}$ being

\begin{equation} \label{Eq23} [a_{d},a_{d}^{\dagger}] =
[a_{g},a_{g}^{\dagger}] = 1. \end{equation}

\noindent These relations lead to

\begin{equation} \label{Sis4}
and \hspace {2cm} \begin{array}{ll}
[{\bf H},(a_{d(g)})^n] = -n (a_{d(g)})^n \hbar\omega \\
\lbrack {\bf H},(a_{d(g)}^{\dagger})^n \rbrack= n(a_{d(g)}^{\dagger})^n
 \hbar\omega.
\end{array}
\end{equation}

In the case of angular deficits neither the $a_{x(y)}$ nor the $a_{d(g)}$
operators can be defined as operators acting on the Hilbert space. The
reason is that they do not respect the periodicity of space, leading states
that respect the periodicity, belonging to the Hilbert space, to states that
do not respect, outside the Hilbert space. The products between them
appearing in (\ref{Eq23}) are nevertheless well defined operators acting
on the Hilbert space since these products do not change the periodicity
in angular variables.  The decomposition of the Hamiltonian in eq.
(\ref{Eq23}) is allowed in the conic space-time once one considers the space of
functions that represents the Hilbert space as embedded in a large space of functions of
arbitrary periodicity. In order to properly define
ladder  operators acting within the Hilbert space it is necessary to define
fractionary powers of the usual creation and annihilation
operators. These highly non-local operators can be constructed for instance as
the infinite series

\begin{equation} \label{series} ({\bf a}_{d(g)}^\dagger)^{\frac{1}{\alpha}} =
\lim_{\epsilon \rightarrow 0}(\epsilon + {\bf a}_{d(g)}^\dagger)^{\frac{1}
{\alpha}} =  \lim_{\epsilon \rightarrow 0}{\sum_{i=0}^{\infty}} C_{i,\frac{1}
{\alpha},\epsilon}({\bf a}_{d(g)}^\dagger)^{i} \end{equation}

\noindent where the regularization parameter $\epsilon$ is to be removed
after summing the series. In this way it is straightforward to
extend the commutation relations (\ref{Sis4}) for

\begin{equation} \label{Sis5}
and \hspace{2cm}\begin{array}{ll}
[{\bf H},(a_{d(g)})^\frac{n}{\alpha}] = -\frac{n}{\alpha}
(a_{d(g)})^\frac{n}{\alpha}
 \hbar\omega \\
\lbrack {\bf H},(a_{d(g)}^{\dagger})^\frac{n}{\alpha} \rbrack=
\frac{n}{\alpha}
(a_{d(g)}^{\dagger})^\frac{n}{\alpha} \hbar\omega.
\end{array}
\end{equation}

\noindent what allows for an interpretation in terms of fractionary quantum
creation and destruction operators. The operator product
$ ({\bf a}_{g}^\dagger {\bf a}_{d}^\dagger)$ also does not change the
periodicity condition and can be in principle defined in the Hilbert space.
Indeed the axial symmetric states not depending on the angular variable
are eigenstates in both the usual and in the conic space cases, being insensitive to
the topological defect of the space-time. These
states are created by applying this operator product on the fundamental state.
We are  thus led to  construct the general basis state vector as

\begin{equation}
  \label{Eq25} |n,n'\rangle^{\alpha}_{g(d)} = ({\bf a}_{g}^\dagger
{\bf a}_{d}^\dagger)^{n}({\bf a}_{g(d)}^\dagger)^{\frac{n'}{\alpha}}
|0,0\rangle. \end{equation}

\noindent Here we avoided the use of the fractionary power operators
 $({\bf a}_{g}^\dagger)^{\frac{1}{\alpha}}$ and
$( {\bf a}_{d}^\dagger)^{\frac{1}{\alpha}} $ at the same time.
Indeed inspection shows that the action of their product on the fundamental
state leads to non-normalizable states. Therefore, all states of the model are
given by linear combinations of

\begin{equation} \label{Sis6}
and \hspace{2cm}  \begin{array}{ll}
|n,n'\rangle^{\alpha}_g & \\
|n,n'\rangle^{\alpha}_d. & \\
\end{array}
\end{equation}

\noindent The energy of the basis states can be calculated, through
relations (\ref{Sis4}) and (\ref{Sis5}) as

\begin{equation} \label{Eq26} E = \hbar \omega ( 2n + \frac{n'}{\alpha} + 1)
.\end{equation}

\noindent It is also straightforward, using eq. (\ref{series}),  to see that
\begin{equation}
  \label{autofuncao}\langle \rho,\phi |0,n'\rangle^{\alpha}_{g(d)} =
\langle \rho,\phi |
({\bf a}_{g(d)}^\dagger)^{\frac{n'}{\alpha}}
|0,0\rangle \end{equation}
\noindent gives the eigenfunctions obtained by the direct solution
of the differential equations. Further applying the differential operator
$({\bf a}_{g}^\dagger
{\bf a}_{d}^\dagger)^{n}$ reproduces all the basis eigenfunctions.

Let us now discuss the Hidden symmetry of the oscillator. It is well know
that angular momentum algebra describes the degeneracy of the 2-D oscillator.
The ``angular momentum'' operators are defined as

\begin{equation}\label{angularmomemtum}
{\bf J}_{\pm}= {\bf a}_{d(g)}^\dagger {\bf a}_{g(d)}
\end{equation}
\noindent
and

\begin{equation}\label{angularmomentum1}
{\bf J}_z= \frac{1}{2} ({\bf a}_{d}^\dagger {\bf a}_{g}- {\bf a}_{g}^\dagger
{\bf a}_{d}).
\end{equation}

\noindent The Casimir operator ${\bf J}^2$ is

\begin{equation}\label{casimir}
{\bf J}^2=\frac{1}{2}({\bf J}_+{\bf J}_-+{\bf J}_-{\bf J}_+)+{\bf J}_z^2=
\frac{1}{2}({\bf N}_g+{\bf N}_d)(
\frac{1}{2}({\bf N}_g+{\bf N}_d)+1),
\end{equation}

\noindent where

\begin{equation}\label{number}
{\bf N}_g+{\bf N}_d= {\bf a}_g^\dagger{\bf a}_g+{\bf a}_d^\dagger{\bf a}_d=\frac{\bf H}{\hbar\omega}-1.
\end{equation}

\noindent This leads to the identification of the $J$ quantum number associated to the
square of the angular momentum with the eigenvalue of $\frac{( { \bf N}_g+{\bf N}_d)}
{2}$

\begin{equation}\label{J}
J=\langle\frac{N_g+N_d}{2}\rangle.
\end{equation}

In the case of angular deficit neither the ${\bf J}_+$ nor the ${\bf J}_-$
operators are defined as operators acting in the Hilbert space of the 2-D
harmonic oscillator spoiling the hidden SU(2) symmetry.
These operators  should be exchanged by ${\bf J}_{\pm}^{\frac{1}{\alpha}}$
to act in the Hilbert space. Nevertheless the composite operators
${\bf J}^2$, ${\bf J}_z$, ${\bf N}_g$ and ${\bf N}_d$ are bona-fide
operators. The relationship expressed by e's. (\ref{casimir}-\ref{J}) are
extended to the deficit angular space case.
 Let us consider then the action of ${\bf N}_g+{\bf N}_d$ on the basis
 states eq. (\ref{Eq25}). Taking $n^\prime=2m$ in that equation it is straightforward
 to see that

 \begin{equation}
 \frac{{\bf N}_g+{\bf N}_d}{2}|n,2m\rangle^\alpha_{g(d)}=(n+\frac{|m|}{\alpha})
 |n,2m\rangle^\alpha_{g(d)}.
               \end{equation}

 \noindent In other words these states have quantum numbers

 \begin{equation}\label{casimirvalue}
 j=n+\frac{|m|}{\alpha}
 \end{equation}

 \noindent This reproduces exactly the form obtained in the
section (2.A) by the resolution of the angular differential equations.

It can be also understood why the operators $(J_\pm )^{1/\alpha}$ do not
generate all states of a multiplet. Since $a^\dagger_{g(d)}$ and $a_{g(d)}$
appear simultaneously with fractional powers in   $(J_\pm )^{1/\alpha}$
they generate non-normalizable functions when applied to the basis states
of eq.(\ref{Eq25}). This is necessary to allow for the change in $j$ value
within the multiplet.

\section{(N+1)-Dimensional generalization}

In order to construct a N-dimensional generalization for Coulomb and
Quantum oscillator problems we consider the metric

\begin{equation} ds^2=dt - (d\rho^2+\frac{1}{\rho^2}d\phi^2+dx_3^2+...+
dx_N^2).
\end{equation}

The variable $\phi$ is assumed to present an angular deficit,
$\phi\rightarrow \phi\alpha$. This space-time generalizes
the cosmic string space-time and the (N-2)-brane is considered in
 $x_1=x_2=0$. We have taken out one generator of the angular momentum
algebra and assumed non trivial boundary conditions for the orbit it
generates in real space breaking thus the  $SO(N)$ symmetry.

In hyper-spherical coordinates the metrics reads:

\begin{equation} g^{\mu\nu} = diag(1, -1, r, r sin\theta ,r sin\theta
sin \phi_{1}, ..., r sin\theta sin \phi_{1}...sin\phi_{N-1}) \end{equation}

\begin{equation} \label{Sis7}
where \hspace{3cm} \begin{array}{ll}
 0 \le r \le \infty  & \\
 0 \le \theta \le \pi & \\
 -\pi \le \phi_{1} \le \pi & \\
 \vdots & \\
 -\pi\alpha \le \phi_{N-1} \le \pi\alpha & \\
\end{array}
\end{equation}

\noindent In this way the N-dimensional Sch\" {o}dinger equation can be written by

\begin{equation} \left[\nabla_{r}^{2} -\frac{\emph{\textbf{L}}^{2}}
{\hbar r^{2}} - \frac{\mu V(r)}{\hbar^{2}}+\frac{2\mu E}{\hbar^{2}} \right]
\Psi(r)=0 \end{equation}

\noindent where

\begin{equation} \nabla_{r}^{2} = \frac{1}{r^{N-1}}\frac{d}{dr}\left
({r^{N-1}}\frac{d}{dr}\right) \end{equation}

\noindent and

\begin{eqnarray} \emph{\textbf{L}}^{2} &=& \frac{1}{sin^{N-2}\theta}
\frac{d}{d\theta}\left(sin^{N-2}{\theta}\frac{d}{d\theta}\right)+
\frac{1}{sin^{2}\theta sin^{N-3}\phi_{1}}
\frac{d}{d\phi_{1}}\left(sin^{N-3}{\phi_{1}}\frac{d}{d\phi_{1}}\right)+ \cdots
 \nonumber \\  & &  \dots+\frac{1}{sin^{2}\theta
sin^{2}\phi_{1} \dots sin^{2}\phi_{N-4}}\frac{1}{sin\phi_{N-3}}
\frac{d}{d\phi_{N-3}}\left(sin{\phi_{N-3}}\frac{d}{d\phi_{N-3}}\right)+
\nonumber \\ & & \hspace{2cm} + \frac{1}{sin^{2}\theta sin^{2}\phi_{1}
\dots sin^{2}\phi_{N-3}}\frac{d^{2}}{d\phi_{N-2}^{2}} \end{eqnarray}

\noindent For V(r) strictly radial we can perform  the separation of variables
method N times and obtain the angular equation

\begin{equation} \emph{\textbf{L}}^{2}Y_{n_{0},\cdots,n_{N-3}}(\theta, \phi_{1},
\cdots, \phi_{N-2})= \ell(\ell+N-2)\hbar^{2}Y_{n_{0},\cdots,n_{N-3}}(\theta, \phi_{1},
\cdots, \phi_{N-2}), \end{equation}

\noindent where  N is the number of dimensions.
Introducing $$k=\sum_{i=0}^{N-3}n_{i}$$  where the integers $n_i$ are separation 
constants the non-trivial boundary condition affects the quantum number 
$\ell$ in the same form as in the three-dimensional case

\begin{equation} l=k+\frac{|m|}{\alpha} \end{equation}

The radial equation will be

\begin{equation} \left[ r^{2}\frac{d^{2}}{dr^{2}} +r(N-1)\frac{d}{dr} +
r^{2} \left(\frac{2 \mu E}{\hbar{2}} -\frac{\mu}{\hbar^{2}}V(r) \right) \right] R(r)
= \ell (\ell +N-2)R(r). \end{equation}

\subsection{N-dimensional solutions for hydrogen atom and quantum harmonic oscillator}

\noindent Particularizing to the Coulomb problem the radial solutions are:

\begin{equation} R(r) = C \left(\frac{r}{r_{0}}\right)^{\ell} e^{-\frac{1}{2} \left(
\frac{r}{r_{0}}\right)}L^{2\ell+N-2}_{i}\left(\frac{r}{kr_{0}}\right) \end{equation}

\noindent where $ k^{2} = -\frac{\epsilon_{0}}{E}$, $i = 1,2...$ and C is a
normalization constant. The energy spectrum is

\begin{equation} E = -\frac{\epsilon_{0}}{\left(i+\ell+\frac{N-3}{2}\right)^{2}}
\end{equation}

For the oscillator problem the radial solutions are:

\begin{equation} R(r) = C \left(\frac{r}{r_{0}}\right)^{\ell} e^{-\frac{1}{2} \left(
\frac{r}{r_{0}}\right)^{2} }L^{\ell +\frac{N-2}{2}}_{i}\left(\frac{r^{2}}{r^{2}_{0}}
\right) \end{equation}

\noindent where C is a normalization constant and and $i = 1,2...$. The energy spectrum is

\begin{equation} E= \hbar\omega \left[\ell +2i+\frac{N-4}{2}\right] \end{equation}

Let us now discuss the relationship between both potential solutions
along the lines discussed for trivial topology by \cite{KM}.

\subsection{Relationship}

With the generalization of the space-time above it is  straightforward to generalize the
mapping  \cite{KM}  of the states of both problems:

\begin{center}
\begin{tabular}{c c c}
& Hydrogen atom  & \hspace{1cm}Harmonic oscillator  \\
\hline
radial Variable  & $\frac{1}{\beta}\frac{r}{r_{0}}$ & $(\frac{r}{{r_{0}^{'}}})^{2}  $  \\
\hline
energy & $\left(\frac{\epsilon_{0}}{E}\right)^{\frac{1}{2}}$&$ \frac{E^{'}}{2\hbar\omega} $      \\
\hline
generalized angular momentum quantum number& $2\ell$ & $ \ell^{'}+\lambda  $  \\
\hline
spatial dimension  &$N$ & $ \frac{N^{'}}{2}-\lambda+1  $  \\
\hline
azimuthal quantum number &  $2\frac{|m|}{\alpha}$ & $\frac{|m'|}{\alpha'}$\\
\hline
angular deficit & $\alpha$ & $\alpha' (\frac{\alpha'}{2})$ \\
\hline
\end{tabular}
\end{center}

\noindent where $\lambda$ allows for a freedom in the mapping.

\

This mapping reveals that the  direct relation of the even states of quantum harmonic
oscillator and all states of hydrogen atom is attenable in the space-time with angular deficit.
It is to be pointed out that the relation can be established with different
angular deficits $\alpha$ and $\alpha\prime)$  and  azimuthal quantum numbers just keeping
$2|m|/\alpha=|m^\prime |/\alpha ^\prime $.

\section{Conclusions}

In this work, we studied the solutions of the Schr\"{o}dinger equation
in  cosmic strings-like space-times with a point in the string acting
as a source for a radial potential. We performed an extension of the spherical harmonics
to the non-trivial space-time. We verified that the global characteristic of
the space-time is present explicitly in the structure of states and energy spectrum.
The deficit angle splits the degeneracies associated with rotational symmetry
in energy spectrum, but the accidental degeneracies are still partially present.

 The extension of the algebraic method of construction of the harmonic oscilator
 states through the introduction of fractionally powered ladder operators
 allowed the discussion of its hidden symmetry. This sheds some light
 on the dependency of the angular momentum Casimir operator values on the angular
 deficit and on the algebraic construction of the angular momentum states.
 In peculiar $\alpha$ values the operators here introduced become local. For instance
 if $\alpha=1/2$ the raising operators become $(a^\dagger)^2$ and the obstruction
 pointed out to  the simultaneous use of left and right raising operators disappears.
 In these cases the presence of the string affects the quantum states simply as a
 supperselection rule.

 Moreover, the remarkable point
raised by this study is that the attempt to allow for the relationship between the coulomb
potential in 3D and the oscillator problem in higher dimensions, if the 3D space
presents conical topology, leads naturally to the construction of the conic
space-times of higher dimensions. In this sense a quantum mechanical
issue is serving as a guide to relate topological space-times in
different dimensions.

The somewhat artificial consideration of the source of the potential point exactly over the string restricts
severally any attempt to  use these results as a means to detect a real cosmic string.

\section*{Acknowledgments}

The authors are grateful to M. F. Alves da Silva for suggesting this study and 
CNPq-Brazil and CAPES-Brazil for partial financial support.


\end{document}